\documentclass[epsfig,11pt]{article}

\usepackage{graphicx}

\begin{document}

\newtheorem{theorem}{Theorem}
\newcommand{\p}{\partial}
\newcommand{\om}{\omega}
\newcommand{\e}{\epsilon}
\renewcommand{\a}{\alpha}
\renewcommand{\b}{\beta}

\title{Impact of Boundary Conditions on Entrainment and Transport in Gravity Currents }

\author{Vena Pearl Bo\~{n}golan-Walsh$^{1}$, Jinqiao Duan$^{1}$, Paul Fischer$^{2}$,\\Tamay \"{O}zg\"{o}kmen$^{3}$, and Traian Iliescu$^{4}$}

\maketitle

\bigskip
\bigskip

{ 1.  Applied Mathematics Department, Illinois Institute of
Technology, Chicago, Illinois\\

2.  Argonne National Laboratory, Darien, Illinois\\

3. RSMAS/MPO,University of Miami, Miami, Florida\\

4.  Department of Mathematics, Virginia Polytechnic Institute and State University, Virginia\\
}

\bigskip
\bigskip

\begin{center}
%Submitted to:\\
To Appear in:\\
{\em Journal of Applied Mathematical Modelling}\\[0.cm]
%{\bf Revised}\\

\bigskip

Dedicated to Professor Guo Youzhong on the occasion of his 70th
birthday.

\end{center}

\date{  }

\newpage
\abstract{

\bigskip

Gravity currents have been studied numerically and experimentally
both in the laboratory and in the ocean.
The question of appropriate boundary conditions is still challenging
for most complex flows.
Gravity currents make no exception - appropriate, physically and
mathematically sound boundary conditions are yet to be found.
This task is further complicated by the technical limitations imposed
by the current oceanographic techniques.

In this paper, we make a first step toward a better understanding of
the impact of boundary conditions on gravity currents.
Specifically, we use direct numerical simulations to investigate the
effect that the popular Neumann, and less popular Dirichlet
boundary conditions on the bottom continental shelf have on the
entrainment and transport of gravity currents.

The finding is that gravity currents under these two different
boundary conditions differ most in the way they transport heat from
the top towards the bottom. This major difference occurs at medium
temperature ranges. Entrainment and transport at high temperatures
also show significant differences.

\bigskip

 }

\newpage

\section{Introduction}

\indent A gravity or density current is the flow of one fluid
within another caused by the temperature (density) difference between the fluids
\cite{Sim82}. Gravity currents flowing down a sloping bottom that
occur in the oceans (e.g., those that flow down the Strait of
Gibraltar and the Denmark Strait) are an important component of
the thermohaline circulation \cite{Cen04}, which in turn is
important in climate and weather studies.

Oceanic gravity currents have been studied in the past, both in the
laboratory and in the ocean. They have been modelled in various
ways, starting from the streamtube models, e.g.,
\cite{Smith,Killworth} to more recent non-hydrostatic Boussinesq
equations~\cite{O3D03, HVBF04, SH97}.

The question of appropriate boundary conditions for complex flows,
such as the gravity currents we consider in this paper, is still
challenging. This paper is a first step in an effort to bridge the
gap between observed gravity currents and the assumptions made in
modeling them. Using ocean observations to develop realistic
boundary conditions for gravity currents is limited by the available
technological means. Thus, the current boundary conditions used in
the numerical simulation of gravity currents are based on physical
intuition and mathematical simplicity and soundness.

The behavior of gravity currents under two different types of
boundary conditions is studied in this paper. Specifically, the
differences between gravity currents flowing with Neumann
(insulation) and Dirichlet (fixed-temperature) bottom boundary
conditions are investigated via direct numerical simulations. Quite
possibly, other boundary conditions, such as Robin, would be more
appropriate. However, given the popularity of Neumann~\cite{ O3D03,
HHU01, OC02, OF04, Hartel} and, to a lesser extent, Dirichlet
boundary conditions in numerical studies~\cite{HKMS97}, we decided
to focus first on these two types of boundary conditions.

%This study takes off from earlier
%studies of overflows that form when cold water from the
%Mediterranean Sea meets the warmer water in the Atlantic~\cite{BK91}.

Bottom Neumann boundary conditions for temperature may be assumed
when the material on the continental shelf is a bad
thermal-conductor (i.e., the current flows over ``insulated'' rock).
Dirichlet boundary conditions may be assumed when the material on
the continental shelf is a good thermal-conductor, making
temperature nearly constant~\cite{Ock99}.

Dirichlet boundary conditions would be appropriate for the initial
transient development of gravity currents. For instance, it is known
that the Red Sea overflow
shuts off in the summer~\cite{MJ97}. Once this gravity current
starts again, one could expect the temperature difference between
the bottom layer and the overflow to have a transient impact on the
mixing near the bottom. Also, such a temperature gradient could
significantly affect the initial neutral buoyancy level, which is of
ultimate interest for large-scale ocean and climate studies.
\"Ozg\"okmen et al.~\cite{OFJ06} found in numerical simulations that
the neutral buoyancy level did not differ significantly from an
analytical estimate, which does not account for mixing, because the
bottom layer properties were isolated from vigorous mixing between
the gravity current and the ambient fluid, and yet determined the
neutral buoyancy level. Thus, any mechanism that can affect the
properties of near bottom fluid is likely to change the neutral
buoyancy level, at least during the initial stages of the
development. This idea will be explored in a future study.

The rest of the paper is organized as follows:
Section~\ref{s_mathematical_setting} presents the mathematical model
used in our numerical study.
Section~\ref{s_numerical_setting} presents the numerical model used
and the model configuration and parameters.
Section~\ref{s_bcs} presents the velocity and temperature boundary
and initial conditions used in the numerical simulation.
Section~\ref{s_numerical_results} presents the numerical investigation
of the effect of Neumann and Dirichlet boundary conditions on the
bottom continental shelf on entrainment and transport in gravity
currents.
Five different metrics are used to this end.
Finally, Section~\ref{s_conclusions} presents a summary of our findings
and directions of future research.

\section{Mathematical Setting}
\label{s_mathematical_setting}

Consider a two-dimensional gravity current flowing downstream,
with x as the horizontal (downstream) direction, and z the
vertical direction (Fig. 1).

The momentum and continuity equations subject to the Boussinesq
approximation can be written as:
\begin{eqnarray}
&& \frac{\partial u}{\partial t}+u\frac{\partial u}{\partial x}
+w\frac{\partial u}{\partial z} =-\frac{1}{\rho_0}\frac{\partial
p}{\partial x}+\nu_h\frac{\partial^2 u}{\partial
x^2}+\nu_v\frac{\partial^2 u}{\partial z^2}\;, \\
&& \frac{\partial w}{\partial t}+u\frac{\partial w}{\partial x}
+w\frac{\partial w}{\partial z} =-\frac{1}{\rho_0}\frac{\partial
p}{\partial z}-g\frac{\rho'}{\rho_0} +\nu_h\frac{\partial^2
w}{\partial x^2} +\nu_v\frac{\partial^2 w}{\partial z^2}\;, \\
&& \frac{\partial u}{\partial x}+ \frac{\partial w}{\partial z}=0 \;,
\end{eqnarray}
where $(x,z)$ are the two spatial dimensions, $t$ the time,
$(u,w)$ the two velocity components, $p$ the pressure,
$g=9.81\,m^2s^{-1}$ the gravitational acceleration, and $\nu_h$ and
$\nu_v$ the viscosities in the horizontal and vertical directions,
here assumed to be constants. The way these viscosities appear in
the simulations is actually via their ratio, $r=\nu_v / \nu_h$.

One can argue that, at the horizontal scale being studied ($10$ km),
we may assume a constant horizontal viscosity, $\nu_h$, since its
controlling factors, model resolution and the speed of the gravity
current (the fastest signal in the system), do not warrant a
variable horizontal viscosity.

At the scale at which these experiments were conducted, which is
similar to that in~\"{O}zg\"{o}kmen and Chassignet~\cite{OC02},
rotation is not considered important either, i.e.,
vertical rigidity is not assumed~\cite{Roisin}. In addition, it is
also known that the vertical diffusion in the ocean is
small~\cite{Ledwell}.
It is precisely the theory that vertical mixing
between gravity currents and the ambient fluid is via
eddy-induced mixing and entrainment, not by diffusion, that led to
the assumption that the vertical viscosity may likewise be taken to
be constant, and assumed to be small, in fact.
The ratio $r=\nu_v / \nu_h$ is kept at $r=0.5*10^{-3}$, under the
range that \"{O}zg\"{o}kmen and Chassignet estimate the results to
be insensitive to the values of the vertical viscosity.

A linear equation of state is used
\begin{eqnarray}
\rho'=-\rho_0\,\alpha\,T\;,
\end{eqnarray}
where $\rho_0$ is the background water density, $\alpha$ the heat
contraction coefficient, and $T$ the temperature deviation from a
background value. The equation for heat transport is
\begin{eqnarray}
\frac{\partial T}{\partial t}+u\frac{\partial T}{\partial x}
+w\frac{\partial T}{\partial z} =\kappa_h\frac{\partial^2
T}{\partial x^2}+ \kappa_v\frac{\partial^2 T}{\partial z^2},
\end{eqnarray}
where $\kappa_h$ and $\kappa_v$ are thermal diffusivities in the
horizontal and vertical directions, respectively.
Nondimensionalizing by
$(x,z)=H\,(x^{*},z^{*}),\;$ $(u,w)=\frac{\nu_h}{H}, \; (u^{*},w^{*}),\;
t=\frac{H^2}{\nu_h}\,t^{*},\;
p=\frac{\rho_0\,\nu_h^2}{H^2}\,p^{*},\;$
\smallskip
$T=\Delta T\,T^{*},\;$
where $H$ is the domain depth and  $\Delta T$ is the
amplitude of the temperature range in the system, and dropping $(*)$,
the equations become
\begin{eqnarray}
&& \frac{\partial u}{\partial t}+u\frac{\partial u}{\partial
x}+w\frac{\partial u}{\partial z} =-\frac{\partial p}{\partial
x}+\frac{\partial^2 u}{\partial x^2} +r\frac{\partial^2
u}{\partial z^2}\;, \label{3d1} \\
&& \frac{\partial w}{\partial t}+u\frac{\partial w}{\partial x}
+w\frac{\partial w}{\partial z} =-\frac{\partial p}{\partial
z}+Ra\,T +\frac{\partial^2 w}{\partial x^2}+ r\,\frac{\partial^2
w}{\partial z^2}\;, \label{3d3} \\
&& \frac{\partial u}{\partial x}+\frac{\partial w}{\partial z}=0\;,
\label{3d4} \\
&& \frac{\partial T}{\partial t}+u\frac{\partial T}{\partial x}
+w\frac{\partial T}{\partial z} =Pr^{-1}\left( \frac{\partial^2
T}{\partial x^2} +r\,\frac{\partial^2 T}{\partial z^2}  \right)
\;,\label{3d5}
\end{eqnarray}
where  $Ra=(g\,\alpha \,\Delta T\,H^3)/\nu_h^2$ is the Rayleigh
number (the ratio of the strengths of buoyancy and viscous forces),
$Pr=\nu_h/\kappa_h $ the Prandtl number (the ratio of viscosity and
thermal diffusivity), and $r=\nu_v/\nu_h=\kappa_v/\kappa_h$ the
ratio of vertical and horizontal diffusivities and viscosities.

We consider $0 < r \leq 1$.
This implies that the viscosity and diffusivity in either direction are
\emph{proportional}, and both are small. This is in line with the
theory that diffusion plays only a small role in the vertical mixing
between gravity currents and the ambient fluid, and the mechanism is
via eddy-induced mixing and entrainment.

\section{Numerical setting}
\label{s_numerical_setting}

\subsection{Numerical model}
\label{ss_numerical_model}

The first consideration taken in passing from the mathematical to
the numerical model was the small amount of physical dissipation.
This calls for an accurate representation of the convective
operator so that the numerical dissipation and dispersion do not
overwhelm the assumed physical effects.

Secondly, since small-scale structures are transported with
minimal physical dissipation, accurate long-time integration is
required. Thus, high-order methods in space and time are needed.
The presence of small-scale structures also implies a need for
significant spatial resolution in supercritical regions, which may
be localized in space.

The numerical method implemented by Nek5000, the software used in
the simulations, is described below:

The spatial discretization of~(\ref{3d1})--(\ref{3d5}) is based on the spectral
element method~\cite{Mad89}, a high-order
weighted residual method based on compatible velocity and pressure
spaces that are free of spurious modes
Locally, the spectral element mesh is structured, with the
solution, data and geometry expressed as sums of Nth-order
Lagrange polynomials on tensor-products of Gauss or Gauss-Lobato
quadrature points. Globally, the mesh is an unstructured array of
K deformed hexahedral elements and can include geometrically
nonconforming elements.
Since the solutions being sought are assumed to be smooth, this
method is expected to give exponential convergence with N,
although it has only $C^{0}$ continuity. The convection operator
exhibits minimal numerical dissipation and dispersion, which is
required in this study.
The spectral element method has been enhanced through the
development of a high-order filter that stabilizes the method for
convection dominated flows but retains spectral accuracy
\cite{Fis01}.

The time advancement is based on second-order semi-implicit
operator splitting methods~\cite{Per93,MaP90}. The
convective term is expressed as a material derivative, and the
resultant form is discretized via a stable backward-difference
formula.

Nek5000 uses an additive overlapping Schwarz method as a
preconditioner, as developed in the spectral element context by
Fischer et al.~\cite{Fis97, Fis00}.
This implementation features fast local solvers that
exploit the tensor-product form, and a parallel coarse-grid solver
that scales to 1000s of processors~\cite{Fis96, Tuf99}.

\subsection{Model configuration and parameters}
\label{ss_model_configuration}

\begin{figure}[t]
\includegraphics[scale=.5]{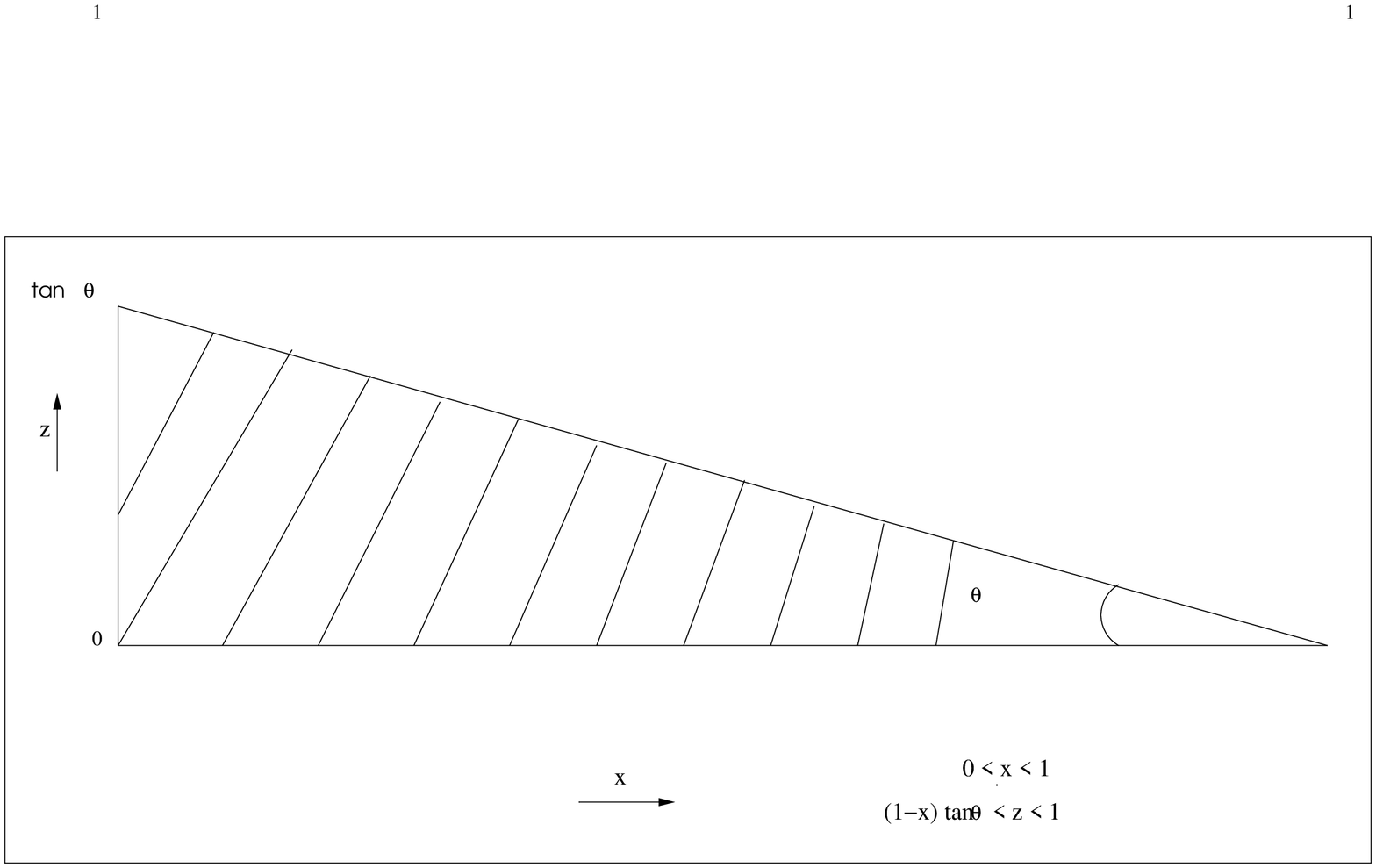}
\caption{Set-up}
\label{figure_1}
\end{figure}

The model domain (see Fig.~\ref{figure_1}) is configured with
a horizontal length of $L_x=10\,km$.
The depth of the water column ranges from $400\,m$ at
$x=0$ to $H=1000\,m$ at $x=10\,km$ over a constant slope. Hence the
slope angle is $\theta=3.5^{\circ}$, which is within the general
range of oceanic overflows, such as the Red Sea overflow entering
the Tadjura Rift~\cite{O3D03}.

Numerical simulations were successfully carried out at a Prandtl
number $Pr = 7$ and a Rayleigh number $Ra = 320 \times 10^{6}$.

The numerical experiments were conducted on a Beowulf Linux cluster
consisting of 9 nodes with 1 Gbps ethernet connectivity. Each node
has dual Athlon 2 GHz processors with 1024 MB of memory.
The 2D simulations take approximately 72 hours (simulated to real time
ratio of $\sim$ 3).

\section{Velocity and temperature boundary and initial conditions}
\label{s_bcs}

\subsection{Temperature boundary conditions}
\label{ss_temperature_bcs}

\paragraph{Inlet}($x=0; \ 0.6 \leq z \leq 1.0$):
The temperature  gradient is one of two agents
driving the flow, the other being the velocity boundary
conditions (see Section~\ref{ss_velocity_bcs}).

\begin{figure}[b]
\includegraphics{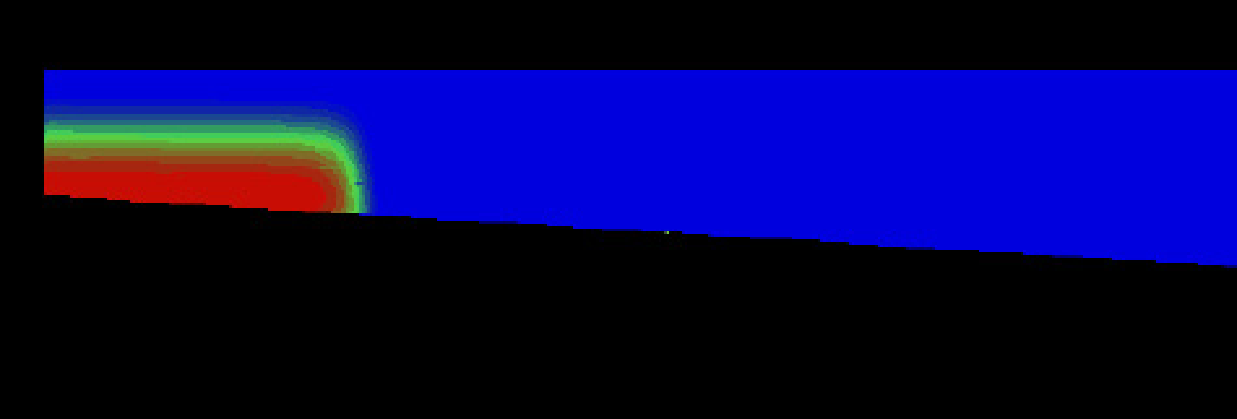}
\caption{Color-coded density plot for temperature at time t=0; red is
coldest, blue is least cold}
\label{figure_2}
\end{figure}

The Dirichlet (fixed-temperature) boundary condition is assumed
at the inlet.
This assumption is simple to work with, although its correctness
ultimately depends on whether or not sea water is an excellent
thermal conductor~\cite{Ock99}.
The boundary condition
$$
T=\frac{1}{2}\left[1-\cos\left( \pi \frac{1-z}{0.4}\right)\right]\;,
$$
and the initial condition
$$
T=\frac{1}{2}\exp(-x^{20})\left[1-\cos\left(\pi \frac{1-z}{0.4}\right)\right]\;,
$$
are used (see Fig.~\ref{figure_2}). Figure~\ref{figure_2} shows the
initial color-coded density plot for temperature (red is lowest
temperature, blue is highest).

\paragraph{Bottom}(the continental shelf):
This is the interface between the gravity current and the
continental shelf. The actual thermal conductivity of the continental
shelf will suggest the proper boundary condition.
Ockendon et al.~\cite{Ock99} prescribe Dirichlet conditions if the boundary
(the continental shelf) is an excellent thermal conductor;
Neumann if the region outside of the gravity current (the
continental shelf) is a very bad thermal conductor; and Robin when
the thermal flux at the boundary is proportional to the difference
between the boundary temperature and some ambient temperature. This last
possibility was not studied in this paper, but could be the topic of
further study, in the light of the results on incremental transport
discussed later in this paper.

Surveys of the literature have not clarified the thermal
conductivity of the continental shelf. The continental shelf is in
the neritic zone, so it would contain a mixture of mineral and
organic materials. In numerical simulations of gravity currents, the
most popular choice is Neumann boundary conditions~\cite{O3D03,
HHU01, OC02, OF04, Hartel}. Fewer numerical studies employ Dirichlet
boundary conditions~\cite{HKMS97}. Other boundary conditions (such
as Robin) might be more appropriate and should be investigated. In
this study, we focus on two different sets of boundary conditions:
Neumann ($\partial T / \partial n = 0$), and Dirichlet ($T = 1$).

Dirichlet boundary conditions would be appropriate for the initial
transient development of gravity currents. For instance, it is known
that the Red Sea overflow
shuts off in the summer~\cite{MJ97}. Once this gravity
current is initiated again, one could expect the temperature difference
between the bottom layer and the overflow to have a transient
impact on the mixing near the bottom. Also, such a temperature
gradient could significantly affect the initial neutral buoyancy level,
which is of ultimate interest for large-scale ocean and climate studies.
\"Ozg\"okmen et al.~\cite{OFJ06} found in numerical simulations that the neutral
buoyancy level did not differ significantly from an analytical estimate,
which does not account for mixing, because the bottom layer properties
were isolated from vigorous mixing between the gravity current and the
ambient fluid, and yet determined the neutral buoyancy level.
Thus, any mechanism that can affect the properties of near bottom fluid
is likely to change the neutral buoyancy level, at least during
the initial stages of the development. This idea will be explored
in a future study.

For Dirichlet boundary conditions, a nondimensional temperature of
one was assumed.

%This would be consistent with the assumed temperature profile of the
%gravity current, with the coldest part at the bottom.

Numerical experiments were conducted with shelf fixed-temperature
values $0 \leq T \leq 1$. $T=0$ had the effect of greatly
speeding-up the gravity current, and not allowing the head to form.
Thus, the current could not entrain ambient water at higher
temperature values, contrary to experimental and oceanic
observations.

\paragraph{Outlet}($x=10; \ 0 \leq z \leq 1$):
Note that the outlet is actually a sea water to
sea water interface, since the region of interest (10 km) is only a
small portion of the ocean's breadth.
The assumed boundary condition here is insulation, i.e.
$\partial T/\partial \mathbf{n}=0$.
(Another possible choice is the ``do-nothing'' boundary
conditions~\cite{Ren97}.)
Since the outlet is far away from
the current for most of its journey down the continental
shelf, the choice of boundary conditions will not influence
significantly our numerical results.

\paragraph{Top}($z = 1$):
This is an air-sea water interface. Presently, insulation ($\partial
T/\partial \mathbf{n}=0$) is assumed. Since air is a bad thermal
conductor, this assumption is justified~\cite{Ock99}.

\subsection{Velocity boundary conditions}
\label{ss_velocity_bcs}

\begin{figure}[t]
\includegraphics[scale=.7]{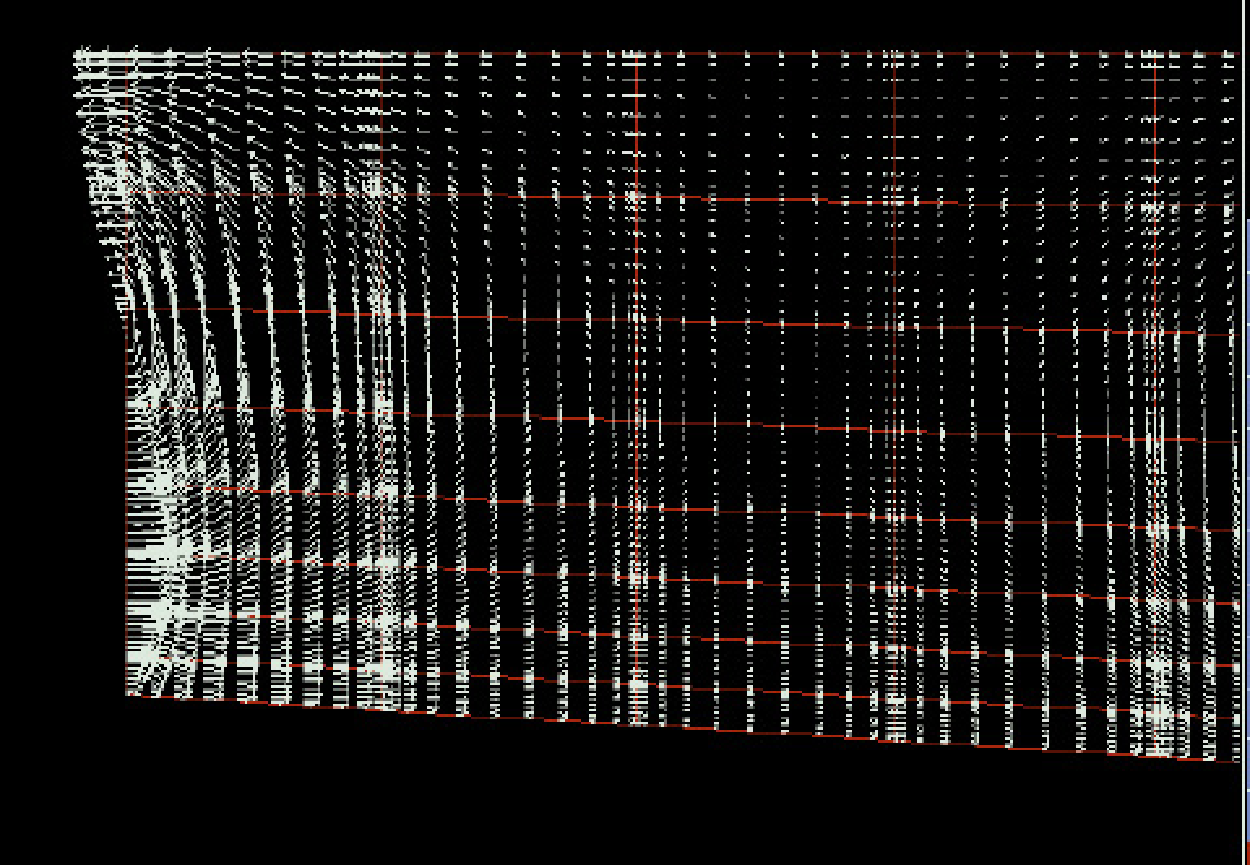}
\caption{Velocity field at the inlet} \label{figure_3}
\end{figure}

\paragraph{Inlet}($x=0; \ 0.6 \leq z \leq 1.0$):
Dirichlet velocity boundary conditions are used at the inlet.
The velocity profile is given by an ``S-shaped''
polynomial function.
The inlet velocity profile is one of
the two forces that drive the flow, the other being the
temperature gradient (see Section~\ref{ss_temperature_bcs}
and Figure~\ref{figure_3}).

\paragraph{Bottom}(the continental shelf):
No-slip velocity boundary conditions ($u = 0, w = 0$)
were used on the bottom.
Previous studies on gravity currents by
\"{O}zg\"{o}kmen et al.~\cite{OC02, O3D03} assume no-slip velocity
boundary conditions on the continental
shelf, which is one of the most popular assumptions for fluids flowing over a
solid boundary. Even in their studies on gravity currents over complex
topography \cite{OF04}, which might constitute ``roughness'' and affect
the log layer, \"{O}zg\"{o}kmen et al. still used the no-slip assumption.
One should note, however, that the question of appropriate boundary
conditions flor fluid flows is still an active area of research.
Other (more appropriate) choices include slip-with-friction boundary
conditions~\cite{Lay99, Lia99, Joh02}.
Given the popularity of Dirichlet boundary conditions in numerical
studies of fluid flows, however, we decided to focus on them first.

Interestingly, H\"artel et al. \cite{Hartel} performed experiments
on gravity currents at high Reynolds numbers for both slip and
no-slip, and did not observe qualitative changes in the flow
structure, although they did observe quantitative Reynolds
number effects.

\paragraph{Outlet}($x=10; \ 0 \leq z \leq 1$):
Free-slip velocity boundary conditions
($\partial w/\partial x=0$,
 $\partial w/\partial y=0$,
 $\partial u/\partial x=0$,
 $\partial u/\partial y=0$)
were used at the outlet.
Another (more appropriate) choice is the ``do-nothing''
boundary conditions~\cite{Ren97}.
Since the outlet is far away from
the current for most of its journey down the continental
shelf, the choice of velocity boundary conditions will not
influence significantly our numerical results.

\paragraph{Top}($z = 1$):
Free-slip velocity boundary conditions were used, similar to those
used at the outlet.

\bigskip

Note that there are no boundary conditions imposed on pressure.
To handle this, Nek5000 uses an operator splitting method, which
initially uses
a ``guess for pressure'', corrects this guess (but velocity is now
non-solenoidal), then corrects the velocity field to impose the
divergence-free velocity field. See Rempfer~\cite{Remp03} for a
discussion of boundary conditions for pressure.

\subsection{Velocity and temperature initial conditions}
\label{ss_initial_conditions}

The initial velocity field is plotted in Fig.~\ref{figure_3}.
The model is entirely driven by the velocity and temperature
(Fig.~\ref{figure_2}) forcing functions at the inlet boundary
($x=0$), and the temperature gradient. The velocity distribution
at this boundary  matches no-slip at the bottom and free-slip at the
top using fourth-order polynomials  such that the depth integrated
mass flux across this boundary is zero.

Other polynomials profiles may be assumed, but care
must by taken to ensure that
\begin{itemize}
\item {the depth-integrated mass flux across the boundary be zero;}
\item {the assumed temperature profile be consistent with the flow
       reversal for the velocity boundary condition; and}
\item {the amplitude of inflow velocity be time-dependent,
       and scaled with the propagation of the gravity current, otherwise
       there will be a recirculation at the inlet (in the case of
       over-estimation) or a thinning-down of the gravity current as it
       flows down-slope, in the case of under-estimation.}
\end{itemize}
These considerations on initial conditions were taken into account
by \"{O}zg\"{o}kmen et al.~\cite{OC02, OF04}.

\section{Numerical results}
\label{s_numerical_results}

Gravity currents evolving with two different sets of boundary
conditions (Neumann/insulation and Dirichlet/fixed temperature), are
studied. Measurements were made of their entrainment, average
temperature, and the heat each transports from the top (warmer
water) towards the bottom (colder water).

\begin{figure}[b]
\includegraphics[scale=.7]{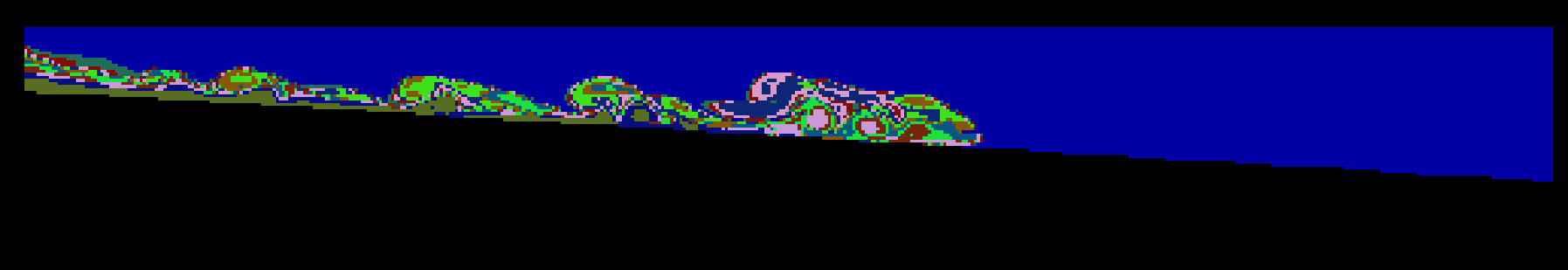}
\caption{ Gravity current with Neumann boundary conditions at time
6375 seconds }
\label{figure_4}
\end{figure}

Figure~\ref{figure_4} shows the gravity current with Neumann boundary
conditions at 6375 seconds (real time). The head and secondary
features are very visible at this point. One can see
the current entraining some of the surrounding fluid via
the Kelvin-Helmholtz instability, which is the main mechanism for
mixing in a non-rotating fluid~\cite{OC02}.

\begin{figure}[t]
\includegraphics[scale=.7]{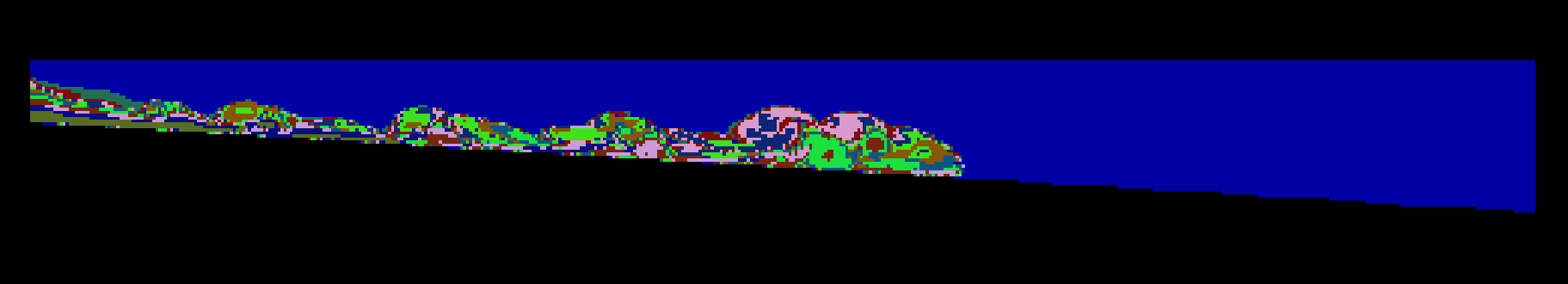}
\caption{ Gravity current with Dirichlet boundary conditions at time
6375 seconds }
\label{figure_5}
\end{figure}

Figure~\ref{figure_5} shows the gravity current with Dirichlet
conditions, also
at 6375 seconds. One can see a difference in the temperature
distributions of the two currents, and it is such differences that
this study tries to quantify.

The differences between the two currents are most notably in the way
they transport heat, and in the way they entrain ambient water. In
order to quantify the differences, five different metrics are used
in
Sections~\ref{ss_entrainment}-\ref{ss_incremental_heat_transport}.
These metrics allow a careful assessment of the effect of Dirichlet
and Neumann boundary conditions on the numerical simulation of
gravity currents.

\subsection{Entrainment $E(t)$}
\label{ss_entrainment}

This metric was earlier defined by Morton et al.~\cite{MTT56}.
The equivalent formulation of \"{O}zg\"{o}kmen et
al.~\cite{O3D03} is used here:
\begin{eqnarray}
E(t) = \frac {\mbox{$h - h_o$}} {\mbox{length}} ,
\end{eqnarray}
where $h$ is the mean thickness of the current from a
fixed reference
point (which is set at 1.28 km), and $h_o$ is the mean thickness of the
tail. The length  of the current is taken from the fixed reference point to
the leading edge of the current. This metric gives us a sense of how ambient
water is entrained into the gravity current.

The tail thickness, $h_o$, is calculated from the flux volumes
passing through the fixed reference point. The mean thickness, $h$, is calculated
from the total volume of
the current, from the fixed reference point to the nose. The difference in volumes
would be accounted for by the
ambient water entrained.

The finding here is that the Dirichlet (fixed-temperature) boundary condition
causes greater entrainment  than Neumann (insulated) boundary condition, by about
$14.21\%$, averaged over the entire travel time down the slope.
This is actually consistent with the weighted temperature metrics (below),
which say that the Neumann boundary conditions yield a colder current.

The maximum difference in $E(t)$ between the two currents
occurs after the head has began to break-down, at around 10,000 seconds
(actual time).

\subsection{Velocity-weighted temperature}
\label{ss_velocity_weighted}

Similar to the passive scalar calculations of Baringer and Price~\cite{BP97a},
the velocity-weighted temperature was calculated as
\begin{eqnarray}
\label{velocity_weighted_temperature} T_v(t) = \frac { \int (T * u
\cdot \textbf{n}) \, dz } {\int  (u \cdot \textbf{n}) \, dz},
\end{eqnarray}
where the integration is from the bottom of the current to its top
(and the current is understood as having nondimensional temperatures
$ 0.0 \leq T \leq 0.8$), $\textbf{n}$ is the unit normal
(in the $x$-direction), and ``*'' denotes multiplication. In this section,
and in the remainder of the paper, we use nondimensional
temperatures.

The finding is that the Neumann boundary condition yields a current that is
$5.61 \%$ colder
than the Dirichlet case, and the difference increases almost
linearly with time.

\subsection{Mean Thickness-weighted Temperature}
\label{ss_mean_thickness}

Analogous to the velocity-weighted temperature~(\ref{velocity_weighted_temperature}),
the mean-thickness weighted temperature is calculated as
\begin{eqnarray}
\label{mean_thickness}
T_h(t) = \frac {\int \int (T*h) \, dx \, dz} {\int\int h \, dx \, dz},
\end{eqnarray}
where the limits of integration are the height ($z$-direction) and
width ($x$-direction) of the current (again, where $ 0.0 \leq T \leq
0.8 $), and ``*'' denotes multiplication. This is another way of
estimating the average temperature for the gravity current at any
given time.

Note that metric~(\ref{mean_thickness}) uses the total volume, the
same volume calculation that is used in calculating $h$ (the mean
thickness of the current).
Hence it is referred to as the mean thickness-weighted temperature.
On the other hand, the velocity-weighted
temperature~(\ref{velocity_weighted_temperature}) uses the flux volumes
instead of the total volume. These flux volumes
are what is used in calculating $h_o$ (the thickness of the tail).

By using metric~(\ref{mean_thickness}), we found that the Neumann
boundary conditions cause the current to be  about $9.36 \%$ colder
than the Dirichlet case, slightly more significant than the difference
reported by metric~(\ref{velocity_weighted_temperature}).

\begin{figure}[b]
\includegraphics[scale=.8]{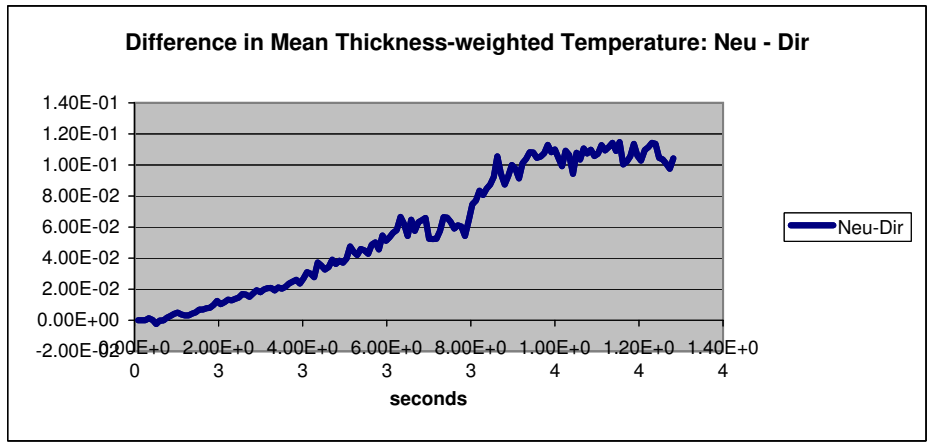}
\caption{ Difference in mean thickness-weighted temperature between
Neumann (Neu) and Dirichlet (Dir) cases}
\label{figure_6}
\end{figure}

Figure~\ref{figure_6} shows the difference in the mean thickness-weighted
temperatures. Just like in Section~\ref{ss_velocity_weighted},
Neumann boundary
conditions always generate a colder current, and the
difference between the average temperatures increases linearly with time.

Both weighted-temperature metrics~(\ref{velocity_weighted_temperature})
and~(\ref{mean_thickness}) give results consistent with the finding for
entrainment in Section~\ref{ss_entrainment}.

\subsection{Total heat transport}
\label{ss_total_heat_transport}

Figure~\ref{figure_2} shows the initial temperature
distribution for both currents. The coldest (red) part is at the
bottom, and the warmest (blue) part at the top.

The total heat transport for each current was calculated as
\begin{eqnarray}
R(t) =   \int (T * u \cdot \textbf{n}) \, dz
\end{eqnarray}
at each time-step. $R(t)$ is then normalized with the length of the
current at that time step.

The finding is that, on average, the current with Neumann boundary conditions
transports more than Dirichlet case, by $9.34 \%$.

\begin{figure}[t]
\includegraphics[scale=1]{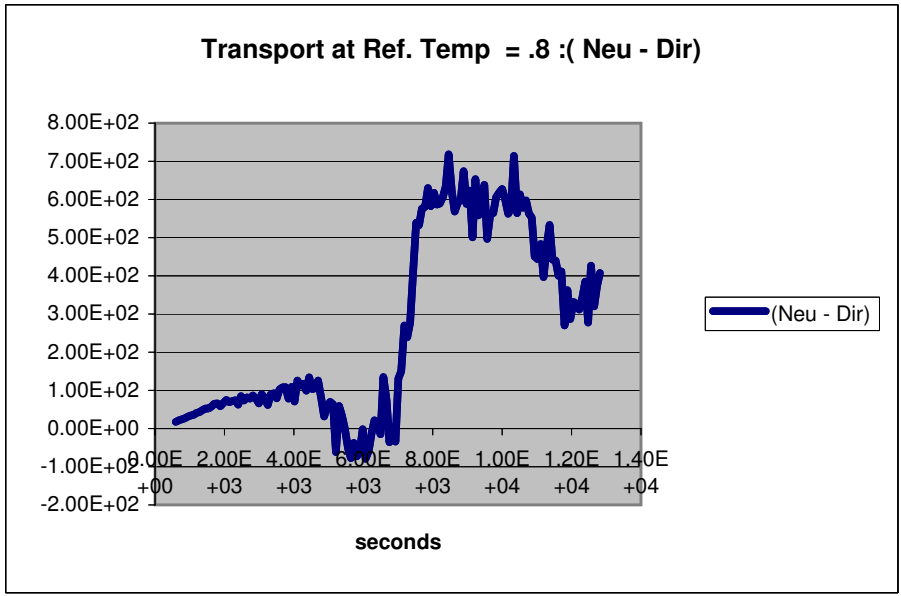}
\caption{ Difference in total temperature transport between Neumann
and Dirichlet cases} \label{figure_7}
\end{figure}

Figure~\ref{figure_7} shows a dramatic rise in the difference in transport
between the two currents at about 6840 seconds (roughly 1/3 of the way down
the domain); the head breaks down at roughly 10,000 seconds (real time).

\subsection{Incremental heat transport}
\label{ss_incremental_heat_transport}

In the same experiment, repeated calls to the flux-volume
calculations are done, changing the reference temperature each time.
This allows the calculation of heat transport for different
reference temperatures.

 The data in
Section~\ref{ss_total_heat_transport} came from the standard
reference temperature of $0.8$, which means that only fluid with
nondimensional temperatures less than or equal to $0.8$ get into the
calculation for $R(t)$ in equation $(13)$ above.

Figure 8 shows the heat transport spectrum plotted for the current
with Neumann boundary conditions, at reference temperatures $0.8,
0.7, 0.5, 0.3$ and $0.1$. Notice that temperature transport
increases as the reference temperatures increases. This is because
the higher the reference temperature, the greater the part of the
current that it encompasses. Indeed, the standard reference
temperature of $0.8$ encompasses all temperatures from $0.0$ to
$0.8$. On the other hand, a reference temperature of $0.7$
encompasses only temperatures from $0.0$ to $0.7$. The Dirichlet
boundary conditions give  a similar spectrum.

The difference in thermal transport between reference temperatures
$0.5$ and $0.7$ is calculated, and this gives an estimate of the way
temperature is transported from the top (warmest part) towards the
bottom (coolest part) of the current. This is equivalent to
calculating $R(t)$ in equation $(13)$ in the previous section, but
only for nondimensional temperatures $ 0.5 \leq T \leq 0.7$.

The same calculation is repeated for the middle temperature range
(from $0.3$ to $0.5$) and low range (from $0.1$ to $0.3$).

The finding here is that, at the middle range, (from $0.3$ to
$0.5$), the Dirichlet case transports more by \textbf{$28.63 \%$}
(see Fig.~\ref{figure_9}). At the high range (from $0.5$ to $0.7$),
Dirichlet continues to transport more, by $11.59\%$, on average.
However, at the low temperature range (from $0.1$ to $0.3$), Neumann
boundary conditions now give a bigger transport, by more than
$7.37\%$. This suggests that there is a major difference in the way
the two currents transport at medium temperatures, and the
difference in the incremental transport increases almost linearly
with time.

\section{Conclusions}
\label{s_conclusions}

Gravity currents with two types of boundary conditions on the bottom
continental shelf, Neumann (insulation) and Dirichlet (fixed
temperature), were investigated. The major finding is that the
incremental temperature transport metric best differentiates the two
currents.

It was noted that the effect is most significant at the medium
temperature range ($0.3$ - $0.5$), where the difference was $28.63
\%$. In this range, the current with Dirichlet boundary conditions
transported more heat.

The other findings (in order of greatest differences) are: The
entrainment $E(t)$ metric also differentiates the two currents
significantly, showing that Dirichlet boundary conditions cause a
current to entrain by about $14.21 \%$ more.  The incremental
temperature transport at the high temperature range showed that the
Dirichlet case transports by $11.59\%$ more. The mean
thickness-weighted temperature showed that Neumann boundary
conditions yield a current that is by about ${9.36\%}$ colder. The
transport at standard reference temperature showed that the Neumann
case transports by $9.34\%$ more.  The incremental transport at the
low temperature range showed that the Neumann boundary conditions
cause about $7.37 \%$ greater transport. Finally, the
velocity-weighted temperature showed that the Neumann case yields a
current that is by about $5.61\%$ colder.

Thus, the observed temperature distribution of the gravity current results from
a complex interaction of temperature being transported from below, and ambient
water being entrained from above.
The present study is just a first step toward a better understanding of this
complex interaction.
In particular, it is found that there can be significant differences in the
entrainment and transport of gravity currents when the most popular Neumann
boundary conditions, or the less popular Dirichlet boundary conditions are
used on the bottom continental shelf.

A better understanding of the impact of both velocity and temperature
boundary conditions on entrainment and transport in gravity currents
is needed.
In particular, slip-with-friction boundary conditions for the velocity
on the bottom continental shelf appear as a more appropriate choice.
Robin boundary conditions for the temperature on the bottom should
also be investigated.
Different outflow velocity and temperature boundary conditions (such as
``do-nothing'') could also yield more physical results, essential in an
accurate long time integration of the gravity current.
All these issues will be investigated in futures studies.

\bigskip

{\bf Acknowledgement.} This work was partly supported by
AFOSR grants F49620-03-1-0243 and FA9550-05-1-0449, and
NSF grants DMS-0203926, DMS-0209309, DMS-0322852, and DMS-0513542.
The authors would like to thank Professors
Xiaofan Li and Dietmar Rempfer for their helpful suggestions,
and the two anonymous reviewers for their insightful
comments and suggestions which improved the paper.

\pagebreak

\pagebreak
\newpage

%% fig 1

%% fig 2

\clearpage
\newpage

% fig 3

%fig 4

\clearpage
\newpage

%fig new 5

%new 6 5

\clearpage
\newpage

%new 7

%new 8
\begin{figure}[b]
\includegraphics[scale=1]{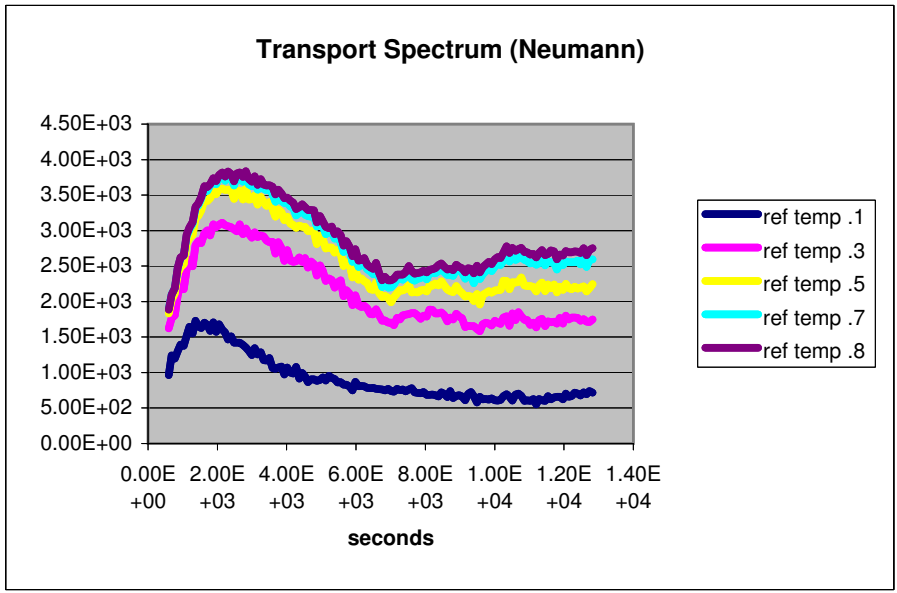}
\caption{ Transport spectrum for Neumann case for various reference
temperatures} \label{figure_8}
\end{figure}

\clearpage
\newpage

%new 9
\begin{figure}[t]
\includegraphics{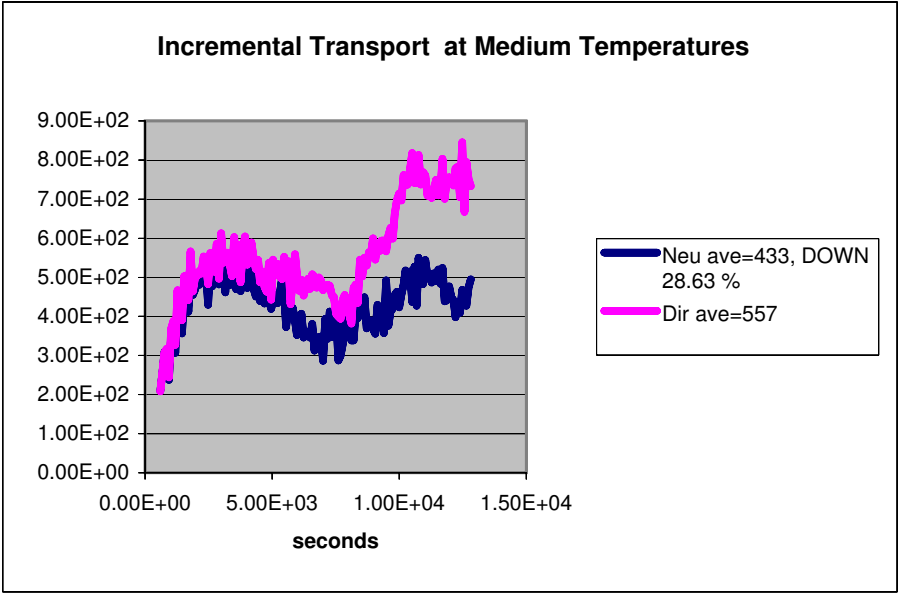}
\caption{ Incremental transport across medium temperature range
$(0.3 - 0.5) $ } \label{figure_9}
\end{figure}

\clearpage

\end{document}